\documentclass[11pt,a4paper]{article}

\usepackage{amsmath,amssymb,wasysym,cite,epsfig,graphicx}

\begin{document}
	
\title{Shadows of rotating Einstein-Maxwell-dilaton black holes surrounded by a plasma}

\author{Javier Bad\'ia$^{1, 2}$\thanks{e-mail: jbadia@iafe.uba.ar} and Ernesto F. Eiroa$^{1}$\thanks{e-mail: eiroa@iafe.uba.ar}\\
	{\small $^1$ Instituto de Astronom\'{\i}a y F\'{\i}sica del Espacio (IAFE, CONICET-UBA),}\\
	{\small Ciudad Universitaria, 1428, Buenos Aires, Argentina}\\
	{\small $^2$ Departamento de F\'{\i}sica, Facultad de Ciencias Exactas y Naturales,} \\ 
	{\small Universidad de Buenos Aires, Ciudad Universitaria Pabell\'on I, 1428, Buenos Aires, Argentina}}
\date{}

\maketitle

\abstract{We consider rotating charged black holes with a scalar dilaton field and surrounded by plasma, with the purpose of studying their shadows. The corresponding metric has been previously obtained in the literature from the static solution by using the Newman-Janis algorithm. Assuming a well known form for the pressureless and nonmagnetized plasma distribution, which is suitable for the separation of the Hamilton-Jacobi equation for light, we derive an expression that determines the shape of the shadow. We present some examples of contours and we analyze their observable properties as functions of the charge and the dilaton coupling. We find that the presence of plasma introduces a dependency on the frequency, with the shadow becoming smaller as the frequency decreases.}

\section{Introduction}

In the last few years, it has been observed for the first time that black holes cast shadows on their surroundings, as predicted by the theory of general relativity \cite{bardeen72,luminet,falcke00}. The Event Horizon Telescope (EHT) Collaboration has produced reconstructed images of both the supermassive black hole M87* at the center of the elliptical galaxy M87 \cite{eht19} as well as Sgr A*, the black hole at the center of our galaxy \cite{eht22}; they show a dark region surrounded by a bright ring of light, which for Sgr A* has a diameter of $\sim 50\, \mu \mathrm{as}$. These observations are consistent with the well-known theoretical scenario in which the trajectories of light rays emitted by the accretion disk of the black hole are deflected by its strong gravitational field, forming a region in a distant observer's sky from which no light arrives. The size and shape of the shadow depend on the various parameters characterizing the black hole and the observer, which for the Kerr solution in general relativity are the mass and the angular momentum of the black hole as well as the inclination angle of the observer. Modified theories of gravity or theories in which general relativity is coupled to additional fields can produce a shadow that is modified with respect to the Kerr shadow, possibly depending on additional parameters. This has motivated the study of black hole shadows as a way of distinguishing Einstein gravity from its alternatives; see Ref. \cite{perlick22} for a review of analytical studies of black hole shadows and Ref. \cite{vagnozzi22} for a thorough testing of alternative geometries against the EHT image of Sgr A*. There have been many publications exploring the present and future possibilities for observing black hole shadows \cite{millimetron,ehi-fromm21}, as well as using them to constrain values of physical parameters \cite{zakharov14} and to test alternative theories of gravity \cite{tests}. Among the many other interesting works in the literature we can mention Refs. \cite{parameters,other,cosmo} concerning shadows in Einstein gravity and Refs. \cite{generales,kumar20,braneworld,otrosalt,otrosalt22} in theories of modified gravity.

One of these alternatives is the Einstein-Maxwell-dilaton (EMD) gravity, in which a scalar field $\phi$ (the dilaton) is coupled to the electromagnetic field $F_{\mu\nu}$ through a term $\exp(-2\lambda \phi)F^2$ in the action, with $\lambda$ a coupling constant. When $\lambda=1$ this theory arises as a low energy limit of string theory, though here we consider a generalization of this limit, where the dilaton is allowed an arbitrary coupling parameter. Due to the presence of the dilaton, charged black holes in string theory do not approach the Reissner-Nordström solution of general relativity at low energies \cite{ghs91}, which in turn can lead to an observable difference between the shadows of charged black holes for both theories \cite{sombrasdilaton}. The static black hole solution in this theory is well known \cite{gibbons88,ghs91}, and its shadow is studied in Ref. \cite{heydari22}. However, finding a rotating solution has proven significantly more difficult; closed form solutions are only known for $\lambda = 0$ (which is simply Einstein-Maxwell theory) and $\lambda = \sqrt{3}$, corresponding to the Kaluza-Klein action \cite{horne92,jensen95}. Therefore, it is necessary to turn to the Newman-Janis algorithm (NJA) \cite{newman65} in order to generate rotating metrics from static solutions---or rather, the so-called \textit{modified Newman-Janis algorithm} \cite{azreg14}, which removes some of the ambiguity present in the original method. We follow Ref. \cite{azreg19}, in which the modified algorithm is used to obtain a rotating solution for arbitrary $\lambda$. It has been shown \cite{lima20} that any metric obtained through the modified NJA admits a separable Hamilton-Jacobi equation for light rays, and thus allows for the analytic calculation of the black hole shadow.

It is expected that astrophysical black holes are surrounded by a plasma medium, and there has been much interest in studying how the properties of the shadow change in the presence of the plasma; see for example Refs. \cite{plasma-review,plasma-varios}. It has also been shown that if the density of a pressureless and nonmagnetized plasma obeys a certain condition \cite{lima20}, then the Hamilton-Jacobi equation for light rays is still separable. The properties of the shadow in this case are chromatic, since the effect of the plasma on the propagation of light depends on the frequency. For low enough frequencies, the black hole develops a ``forbidden region'' which light cannot penetrate, leading to a dramatic decrease of the shadow size \cite{perlick17}. In this work, we arrive at an expression for the contour of the shadows of the rotating black holes obtained from the static solutions of EMD gravity through the modified NJA and surrounded by a plasma obeying the separability condition. We then adopt a simple plasma model and we present the shadow and its geometric properties for various values of the coupling parameter of the theory and the photon frequency, as well as the angular momentum and charge of the black hole. The paper is organized as follows: in Sec. \ref{sec:teoria}, we briefly present Einstein-Maxwell-dilaton theory and discuss its static and rotating black hole solutions. In Sec. \ref{sec:hj}, we introduce the Hamilton-Jacobi equation for light rays in a plasma and find the contour of the black hole shadow, of which we show some examples in Sec. \ref{sec:ex}. Finally, in Sec. \ref{sec:conc} we conclude and discuss our results. Throughout this work we adopt units such that $G = c = \hbar = 1$.

\section{Einstein-Maxwell-dilaton gravity}\label{sec:teoria}

We consider the theory defined by the action \cite{ghs91,gibbons88}
\begin{equation}\label{accion}
	S = \int d^4x \sqrt{-g} \left(R - 2 (\nabla \phi)^2 - e^{-2\lambda \phi} F_{\mu\nu} F^{\mu\nu} \right),
\end{equation}
where $R$ is the scalar curvature associated to the metric tensor $g_{\mu\nu}$ and $\lambda$ is an arbitrary coupling parameter between the electromagnetic field tensor $F_{\mu\nu}$ and the dilaton field $\phi$. Note that changing the sign of $\lambda$ is equivalent to changing the sign of $\phi$, so we can take $\lambda \geq 0$ without loss of generality. When $\lambda = 0$ the action reduces, up to an unimportant overall factor of $1/16\pi$, to the usual Einstein-Maxwell action together with a minimally coupled scalar field. As mentioned before, this action is part of the low-energy limit of string theory when $\lambda=1$. The field equations resulting from Eq. \eqref{accion} read
\begin{equation}\label{eom-f}
	\nabla _{\mu }\left( e^{-2\lambda \phi }F^{\mu \nu }\right) = 0,
\end{equation}
\begin{equation}\label{eom-phi}
	\nabla ^{2}\phi + \frac{\lambda }{2}e^{-2\lambda \phi }F_{\mu \nu }F^{\mu\nu} = 0,
\end{equation}
\begin{equation}\label{eom-g}
	R_{\mu \nu }=2\nabla _{\mu }\phi \nabla _{\nu }\phi +2e^{-2\lambda \phi
	} \left( F_{\mu \alpha }F_{\nu }^{\; \alpha }-\frac{1}{4}g_{\mu \nu
	}F_{\alpha \beta }F^{\alpha \beta } \right) .
\end{equation}

\subsection{Static solution}

The static and spherically symmetric solution to EMD gravity with an arbitrary coupling parameter $\lambda$ has the form\cite{ghs91,gibbons88}
\begin{equation}\label{metrica-estatica}
	ds^2 = -f(r) dt^2 + \frac{dr^2}{f(r)} + h(r) d\Omega^2,
\end{equation}
with
\begin{equation}\label{met-f}
	f(r) = \left(1 - \frac{r_1}{r}\right) \left(1 - \frac{r_2}{r}\right)^{(1-\lambda^2)/(1+\lambda^2)}
\end{equation}
and
\begin{equation}\label{met-h}
	h(r) = r^2 \left(1-\frac{r_2}{r}\right)^{2\lambda^2/(1+\lambda^2)},
\end{equation}
where $r_1$ and $r_2$ are two parameters related to the mass $M$ and charge $Q$ of the black hole by
\begin{equation}\label{masa}
	M = \frac{r_1}{2} + \left(\frac{1-\lambda^2}{1+\lambda^2}\right) \frac{r_2}{2}
\end{equation}
and
\begin{equation}\label{carga}
	Q^2 = \frac{r_1r_2}{1+\lambda^2}.
\end{equation}
The dilaton and the Maxwell fields are given by
\begin{equation}
	e^{2\phi} = \left(1 - \frac{r_2}{r}\right)^{2\lambda/(1+\lambda^2)}
\end{equation}
and
\begin{equation}
	F_{tr} = \frac{Q}{r^2}.
\end{equation}
Equations \eqref{masa} and \eqref{carga} can be inverted to give the radii $r_1$ and $r_2$ in terms of the mass and charge:
\begin{gather}
	r_1 = M + \sqrt{M^2 - (1-\lambda^2)Q^2} \label{r1} \\
	r_2 = \frac{1+\lambda^2}{1-\lambda^2} \left(M - \sqrt{M^2 - (1-\lambda^2)Q^2}\right); \label{r2}
\end{gather}
these equations are quadratic, and the signs have been chosen to give positive solutions. Note that the radii are real, and thus the metric \eqref{metrica-estatica} is well defined, only if $(1-\lambda^2)Q^2 \leq M^2$. This condition is automatically satisfied if $\lambda \geq 1$, but if $\lambda < 1$ it places an upper limit
\begin{equation}\label{q-existe}
	Q^2 \leq \frac{1}{1-\lambda^2}M^2
\end{equation}
on the charge. If the above condition is met, the spacetime may still contain a naked singularity. For $\lambda = 0$, the solution reduces to the Reissner-Nordström metric of general relativity, which has a pair of horizons at $r_{\pm} = r_{1,2}$ and a point singularity at $r=0$. For any $\lambda > 0$ the horizons are still located at $r_{\pm} = r_{1,2}$ but the geometry at $r=r_2$ becomes singular, so we demand that $r_1 > r_2$ in order to avoid a naked singularity \cite{ghs91}. In terms of the charge and mass, this translates into the condition
\begin{equation}
	Q^2 \leq (1+\lambda^2)M^2
\end{equation}
for an event horizon to exist.

\subsection{Rotating EMD black holes}

As mentioned above, rotating solutions to EMD gravity are only known in closed form for $\lambda = 0$ and $\lambda = \sqrt{3}$ \cite{horne92}; the former case is the Kerr-Newman solution, while the latter is the rotating black hole in Kaluza-Klein theory. The Newman-Janis algorithm provides a way to generate rotating metrics from a static ``seed'' metric through a complexification of the coordinates; it was originally used to show how the Kerr metric can be obtained from the Schwarzschild metric and to subsequently produce for the first time the Kerr-Newman solution of general relativity coupled to Maxwell electrodynamics \cite{newman65}. However, the algorithm has two drawbacks. The first one is that it requires one to guess the appropriate complexification of the metric functions \cite{azreg14}, and no prescription or reasoning is given. The other drawback is that, outside general relativity, the metric produced by the algorithm will not satisfy the same field equations as the seed metric. In general, it requires a modified energy-momentum tensor with respect to the original spacetime, usually with the addition of extra fluids or fields \cite{hansen13,beltracchi21}. In fact, applying the NJA to the static solution \eqref{metrica-estatica} with $\lambda = 1$ produces the previously found Kerr-Sen metric \cite{yazadjiev00,sen92}, which is not a solution of the equations of motion \eqref{eom-f}-\eqref{eom-g} unless an extra field, the axion, is added to the action. The modified NJA \cite{azreg14} is an alternative to overcome these problems, in which no guesswork is necessary but instead an overall function multiplying the metric is left undetermined; physical arguments can help to provide a criterion for choosing a specific function. It has been adopted in many articles appearing in the literature in recent years, see e.g. Ref. \cite{lima20} and references therein. In what follows, we will use the results obtained in Ref. \cite{azreg19}, where the modified NJA is used to produce a rotating black hole metric starting from the static seed solution \eqref{metrica-estatica} for arbitrary values of $\lambda$. The line element in Boyer-Lindquist coordinates \cite{azreg19} is given by\footnote{We only consider in this work the normal black holes introduced in Ref. \cite{azreg19}, in which phantom black holes are also studied.}
\begin{equation}\label{metrica-rot}
	ds^2 = - \frac{H \Delta}{\Sigma} dt^2 + \frac{\Sigma \sin^2\theta}{H} \left(d\varphi - \frac{a\sigma}{\Sigma} dt \right)^2 + \frac{H}{\Delta} dr^2 + H d\theta^2,
\end{equation}
where $a = J/M$ is the angular momentum per unit mass,
\begin{gather}
	H = h + a^2\cos^2\theta, \label{h-funcion} \\
	\Delta = fh + a^2 = r^2 - (r_1+r_2)r + r_1 r_2 + a^2, \\
	\sigma = h(1-f), \\
	\Sigma = (h+a^2)^2 - a^2 \Delta \sin^2\theta,
\end{gather}
and the functions $f(r)$ and $h(r)$ are as in Eqs. \eqref{met-f} and \eqref{met-h}. 
In the derivation of this metric, the overall multiplying function was chosen so that to have a null cross term of the Einstein tensor \cite{azreg19}, that is $G_{r\theta}=0$. As a consequence, it is a physically acceptable solution of the field equations, because the energy-momentum tensor can be written in the form \cite{azreg14}
\begin{equation}
    T^{\mu\nu}= \epsilon e^\mu_t e^\nu_t + p_r e^\mu_r e^\nu_r + p_\theta e^\mu_\theta e^\nu_\theta + p_\varphi e^\mu_\varphi e^\nu_\varphi,
\end{equation}
where $(e_t, e_r, e_\theta, e_\varphi)$ is an orthonormal tetrad for which $e_r$ and $e_\theta$ are proportional to the $\partial_r$ and $\partial_\theta$ basis vectors. This means that the source term $T_{\mu\nu}$ can be interpreted as an imperfect fluid rotating about the $z$ axis \cite{azreg14}. 
This geometry has two horizons located at the roots of $\Delta(r)$, given by
\begin{equation}
	r_\pm = \frac{r_1+r_2 \pm \sqrt{(r_1-r_2)^2 - 4a^2}}{2}.
\end{equation}
For $\lambda = 0$, the EMD solution reduces to the Kerr-Newman  geometry of general relativity, with two horizons at $r_{\pm}= M \pm \sqrt{M^2-a^2-Q^2}$ and a ring shaped singularity located at $r=0$ and $\theta =\pi/2$. When $\lambda >0$, the location of the singularity is more complicated than in the static case now depending, besides $\lambda$, $M$, and $Q$, also on $a$ and $\theta$ \cite{azreg19}. In order that the horizons exist and to avoid having a naked singularity, we require
\begin{equation}
	r_1 - r_2 \geq 2|a|
\end{equation}
instead of $r_1 - r_2 > 0$ as in the static case. Rewriting this condition in terms of the mass and charge using Eqs. \eqref{r1} and \eqref{r2} we arrive at
\begin{equation}\label{q-extremal}
	Q^2 \leq (M-|a|) \left[(1+\lambda^2)M + (1-\lambda^2)|a|\right].
\end{equation}
For $\lambda \leq 1$ this can only be satisfied if $|a|/M \leq 1$, while for $\lambda > 1$ there is a second branch of solutions with $|a|/M \geq (\lambda^2 + 1)/(\lambda^2 - 1)$, in addition to those with $|a|/M \leq 1$; we will not consider these higher values of $|a|$, since they have not been observed in astrophysical black holes and the two spaces of solutions are disconnected. For further details about the rotating EMD spacetime, see Ref. \cite{azreg19}.

\section{Black hole shadow in a plasma environment}\label{sec:hj}

We consider the simple case of a cold (i.e., pressureless) and nonmagnetized plasma, in which the motion of light is described by the Hamiltonian \cite{perlick-libro}
\begin{equation}\label{hamiltoniano}
	\mathcal{H}(x,p) = \frac12 \left( g^{\mu\nu}(x) p_\mu p_\nu + \omega_p^2(x) \right)
\end{equation}
where $\omega_p$ is the plasma electron frequency, given in terms of the electron density $N_e(x)$ by
\begin{equation}
	\omega_p^2 = \frac{4\pi e^2}{m_e} N_e,
\end{equation}
with $e$ and $m_e$ the electron charge and mass, respectively. Equivalently, one can define a refractive index $n$ depending on the photon frequency $\omega$ \cite{synge60} by
\begin{equation}
	n^2 = 1 - \frac{\omega_p^2}{\omega^2}.
\end{equation}
Light rays correspond to the solutions of the Hamilton equations with $\mathcal{H}=0$. The metric \eqref{metrica-rot}, being stationary and axisymmetric, is independent of the coordinates $t$ and $\varphi$, and we also assume that the same is true for the plasma frequency $\omega_p(x)$. We then immediately have two conserved quantities $\omega_0 \equiv - p_t$ and $p_\varphi$ along photon trajectories; since the metric is asymptotically flat, $\omega_0$ is the frequency of the photon at infinity. In flat spacetime, light cannot propagate through a plasma if its frequency is low enough; similarly, it can be shown \cite{perlick17,badia21} that for the Hamiltonian \eqref{hamiltoniano} the condition
\begin{equation}\label{prohibida}
	\omega_0^2 \geq - g_{tt} \omega_p^2(r,\theta)
\end{equation}
should be satisfied to allow light with frequency $\omega_0$ to exist at a given spacetime point.

The standard method to integrate the equations of motion was first introduced by Carter \cite{carter68} for the Kerr metric, and it involves finding an additional constant of motion by separating variables in the Hamilton-Jacobi equation
\begin{equation}\label{hj}
	\mathcal{H}\left(x, \frac{\partial S}{\partial x}\right) = 0.
\end{equation}
It was later extended to more general scenarios; the conditions for the equation to be separable in an arbitrary stationary and axisymmetric spacetime with a nonmagnetized plasma were found in Ref. \cite{bezdekova22}. More importantly for this work, it was also previously shown \cite{lima20} that the Hamilton-Jacobi equation is always separable for a metric obtained through the modified Newman-Janis algorithm as long as the plasma frequency has the form
\begin{equation}\label{plasma-sep}
	\omega_p^2 = \frac{f_r(r) + f_\theta(\theta)}{H},
\end{equation}
where $H$ is the metric function \eqref{h-funcion} and $f_r$ and $f_\theta$ are functions of their respective coordinates.. Substituting the inverse metric and the plasma frequency into the Hamiltonian \eqref{hamiltoniano} and proposing an \textit{ansatz}
\begin{equation}
	S = - \omega_0 t + p_\varphi \varphi + S_r(r) + S_\theta(\theta)
\end{equation}
for the action, after some algebra we arrive at the equality
\begin{equation}
	(S_\theta')^2 + \left(a \omega_0 \sin\theta - \frac{p_\varphi}{\sin\theta}\right)^2 + f_\theta = \frac{1}{\Delta} \left[\omega_0(h+a^2) - a p_\varphi\right]^2 - \Delta (S_r')^2 - f_r.
\end{equation}
Since the left-hand side is only a function of $\theta$ and the right-hand side only a function of $r$, both sides must be equal to a constant $K$, known as the Carter constant \cite{carter68}. Putting together the expressions $p_\mu = \partial S / \partial x^\mu$ for the momenta and $\dot{x}^\mu = g^{\mu\nu} p_\nu$ for the velocities, where a dot denotes a derivative with respect to an affine parameter, we can bring the equations of motion to first order:
\begin{gather}
	H \dot{t} = \frac{h+a^2}{\Delta} P(r) - a^2 \sin^2\theta\, \omega_0 + a p_\varphi, \\
	H \dot{\varphi} = \frac{a}{\Delta} P(r) - a \omega_0 + \frac{p_\varphi}{\sin^2\theta}, \\
	(H \dot{r})^2 = R(r), \\
	(H \dot{\theta})^2 = \Theta(\theta), \label{theta-punto}
\end{gather}
where
\begin{gather}
	R(r) = P(r)^2 - \Delta (K + f_r), \\
	\Theta(\theta) = K - \left(a \omega_0 \sin\theta - \frac{p_\varphi}{\sin\theta}\right)^2 - f_\theta, \label{theta} \\
	P(r) = \omega_0 (h+a^2) - a p_\varphi.
\end{gather}
It is straightforward to verify that our results agree with those derived in Refs. \cite{lima20,bezdekova22}.

Of particular interest among the possible trajectories are the spherical photon orbits: solutions with constant $r$, which satisfy $R(r) = R'(r) = 0$. The black hole shadow is defined as the set of directions in an observer's sky which, when continued into the past along light rays, intersect the event horizon. The trajectories that make up its contour are asymptotic to the unstable spherical photon orbits, and therefore have the same conserved quantities as them. The equations $R(r) = R'(r) = 0$ are quadratic in $p_\varphi$ and $K$, so it is possible to solve them analytically as parametric functions of $r$; the solutions, also called the critical values of the constants of motion, are
\begin{gather}
	\frac{a p_\varphi}{\omega_0} = h + a^2 - \frac{\Delta h'}{\Delta'} \left(1 \pm \sqrt{1 - \frac{\Delta' f_r'}{\omega_0^2 h'^2}}\right) \label{pphi-c}, \\
	K = \frac{\Delta \omega_0^2 h'^2}{\Delta'^2} \left(1 \pm \sqrt{1 - \frac{\Delta' f_r'}{\omega_0^2 h'^2}}\right)^2 - f_r. \label{K-c}
\end{gather}
It is straightforward to adapt the argument given in Ref. \cite{perlick17}, that only the plus sign in these solutions is physically relevant, under the condition that the plasma frequency has $f_\theta \geq 0$ and $f_r(r) = C r^k$, with $C \geq 0$ and $0 \leq k \leq 2$. The particular case of plasma that we will consider below does satisfy this condition, so we will take the plus sign in Eqs. \eqref{pphi-c} and \eqref{K-c}. On any trajectory, Eq. \eqref{theta-punto} implies that $\Theta$ must be nonnegative. If for a given $r$ the critical values of $p_\varphi$ and $K$ are substituted into the definition \eqref{theta} of $\Theta$, the inequality $\Theta(\theta) \geq 0$ gives the range of $\theta$ for the chosen photon orbit. In particular, the range of radii at which spherical orbits exist is given by those $r$ for which, after substituting the critical conserved quantities, the inequality $\Theta(\theta) \geq 0$ has solutions. The set of all points through which spherical photon orbits pass is called the photon region.

As explained above, the contour of the black hole shadow as seen by a distant observer consists of light rays which asymptotically approach the spherical photon orbits, and thus share their constants of motion. A given $\omega_0$ and a pair $(p_\varphi, K)$ satisfying Eqs. \eqref{pphi-c} and \eqref{K-c} describe a single photon orbit, and the outgoing light ray with the same conserved quantities corresponds to one direction in the sky of the observer. The set of these directions as $r$ ranges over the photon region is the contour of the shadow. To relate the conserved quantities to directions in the sky, we take the observer to be stationary at an inclination angle $\theta = \theta_\text{o}$ and at a very large distance $r_\text{o}$ from the black hole, taking advantage of the fact that the spacetime is asymptotically flat, and use the orthonormal tetrad
\begin{gather}
	\mathbf{e}_{\hat{t}} = \partial_t, \\
	\mathbf{e}_{\hat{r}} = \partial_r, \\
	\mathbf{e}_{\hat{\theta}} = \frac{1}{r_\text{o}} \partial_\theta, \\
	\mathbf{e}_{\hat{\varphi}} = \frac{1}{r_\text{o} \sin\theta_\text{o}} \partial_\varphi,
\end{gather}
with the tetrad components of the four-momentum of a photon given in terms of the coordinate components by
\begin{gather}
	p^{\hat{t}} = \omega_0, \\
	p^{\hat{r}} = p_r, \\
	p^{\hat{\theta}} = \frac{p_\theta}{r_\text{o}}, \\
	p^{\hat{\varphi}} = \frac{p_\varphi}{r_\text{o} \sin\theta_\text{o}}.
\end{gather}
We additionally assume that the plasma frequency goes to zero at infinity, so that photons propagate in a vacuum when they arrive at the observer. We can then define the celestial coordinates
\begin{gather}
	\alpha = - r_\text{o} \frac{p^{\hat{\varphi}}}{p^{\hat{t}}} \bigg|_{r_\text{o} \to \infty}, \\
	\beta = - r_\text{o} \frac{p^{\hat{\theta}}}{p^{\hat{t}}} \bigg|_{r_\text{o} \to \infty},
\end{gather}
where $\alpha$ measures distances perpendicular to the spin of the black hole, while $\beta$ is parallel to it; the origin of the coordinates corresponds to the optical axis. Finally, using $p_\theta = \partial S / \partial \theta$ to write $p^{\hat{\theta}}$ in terms of the conserved quantities, we arrive at the expressions
\begin{gather}
	\alpha = - \frac{p_\varphi}{\omega_0 \sin\theta_\text{o}}, \label{alfa}\\
	\beta = \pm \frac{1}{\omega_0} \sqrt{K - \left(a \omega_0 \sin \theta_\text{o} - \frac{p_\varphi}{\sin\theta_\text{o}}\right)^2 - f_\theta(\theta_\text{o})}. \label{beta}
\end{gather}
Not all spherical photon orbits actually correspond to directions in the sky if the observer is not equatorial: the constants $p_\varphi$ and $K$ must be such that the square root in Eq. \eqref{beta} is real. For a given $\omega_0$, taking the critical constants of motion \eqref{pphi-c} and \eqref{K-c} as functions of $r$ and replacing them into Eqs. \eqref{alfa} and \eqref{beta} gives a parametric curve $(\alpha(r), \beta(r))$ tracing the contour of the shadow. The range of $r$ is limited by the values $r_\pm$ for which $\beta(r_\pm) = 0$. Note that we must include both signs in the expression for $\beta$, corresponding to the upper and lower halves of the shadow.

\section{Shadows and observables}\label{sec:ex}

\begin{figure}
	\centering
	\includegraphics[width=\linewidth]{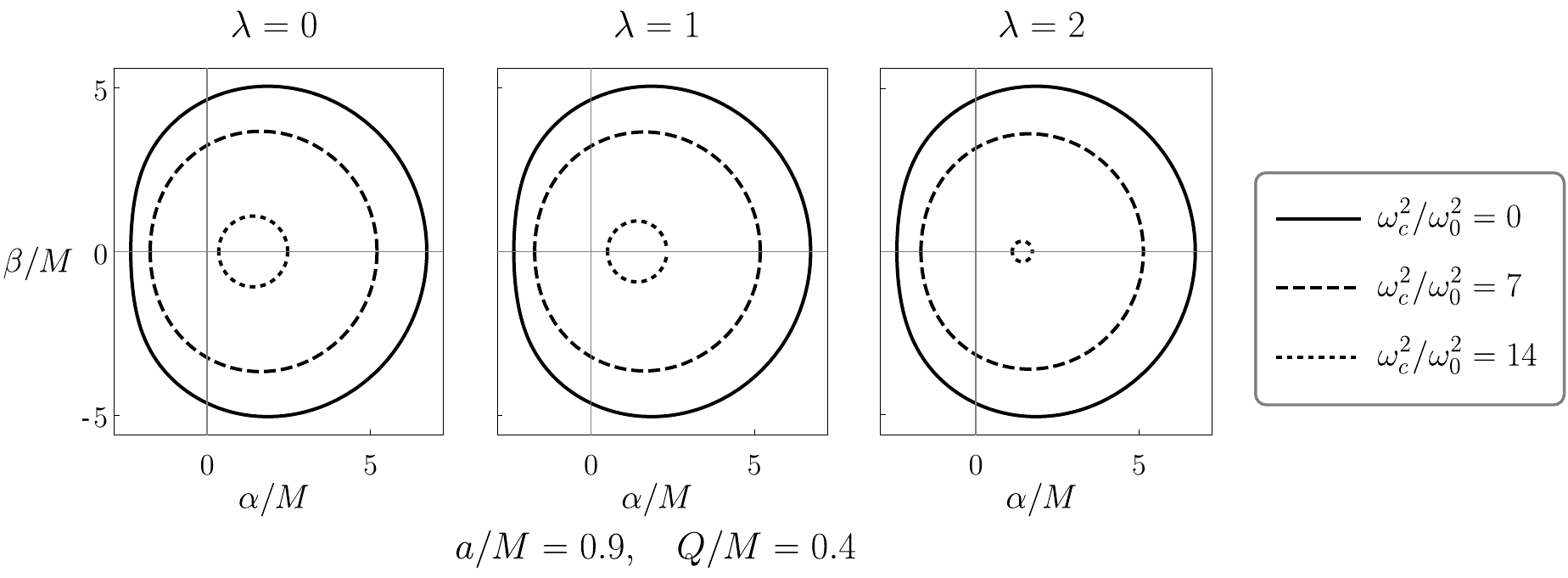}
	\caption{Shadow of a rotating EMD black hole with $a/M = 0.9$ and $Q/M = 0.4$ for an equatorial observer.}
	\label{fig:sombras1}
\end{figure}

\begin{figure}
	\centering
	\includegraphics[width=0.3\linewidth]{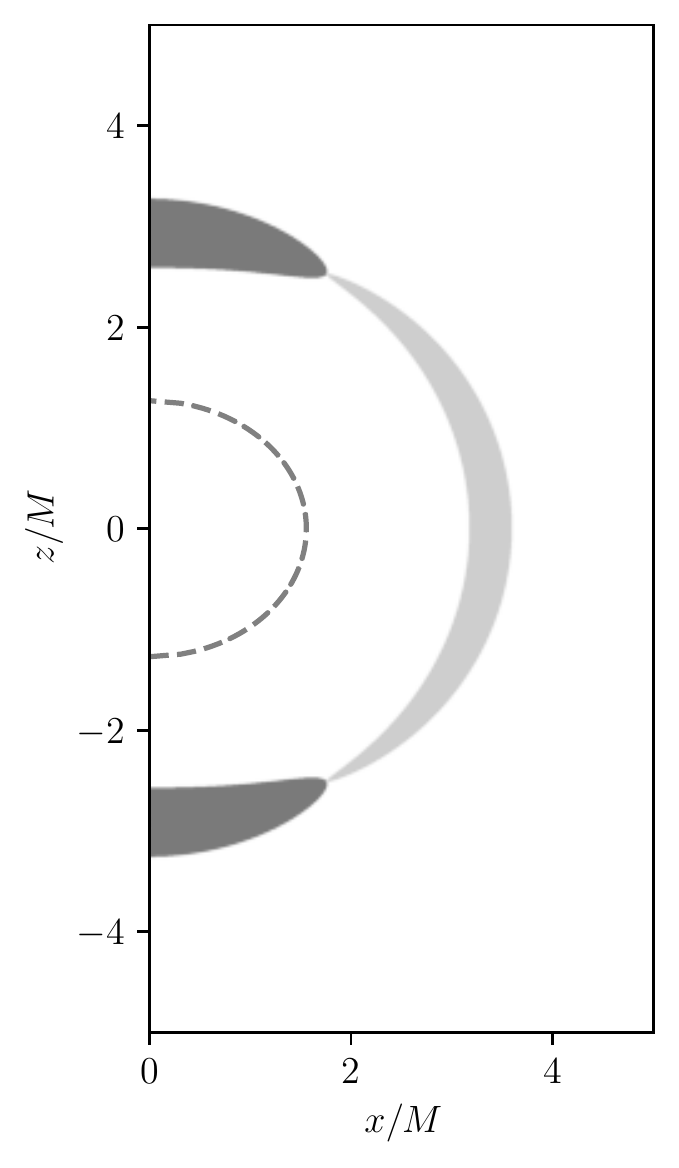}
	\caption{Photon region (light gray) and forbidden region (dark gray) around a rotating EMD black hole with $a/M = 0.9$, $\lambda = 1$, and $Q/M = 0.4$, for photons with $\omega_c^2/\omega_0^2 = 14$. The size of the forbidden region increases rapidly as the frequency decreases.}
	\label{fig:regiones}
\end{figure}

\begin{figure}
	\centering
	\includegraphics[width=\linewidth]{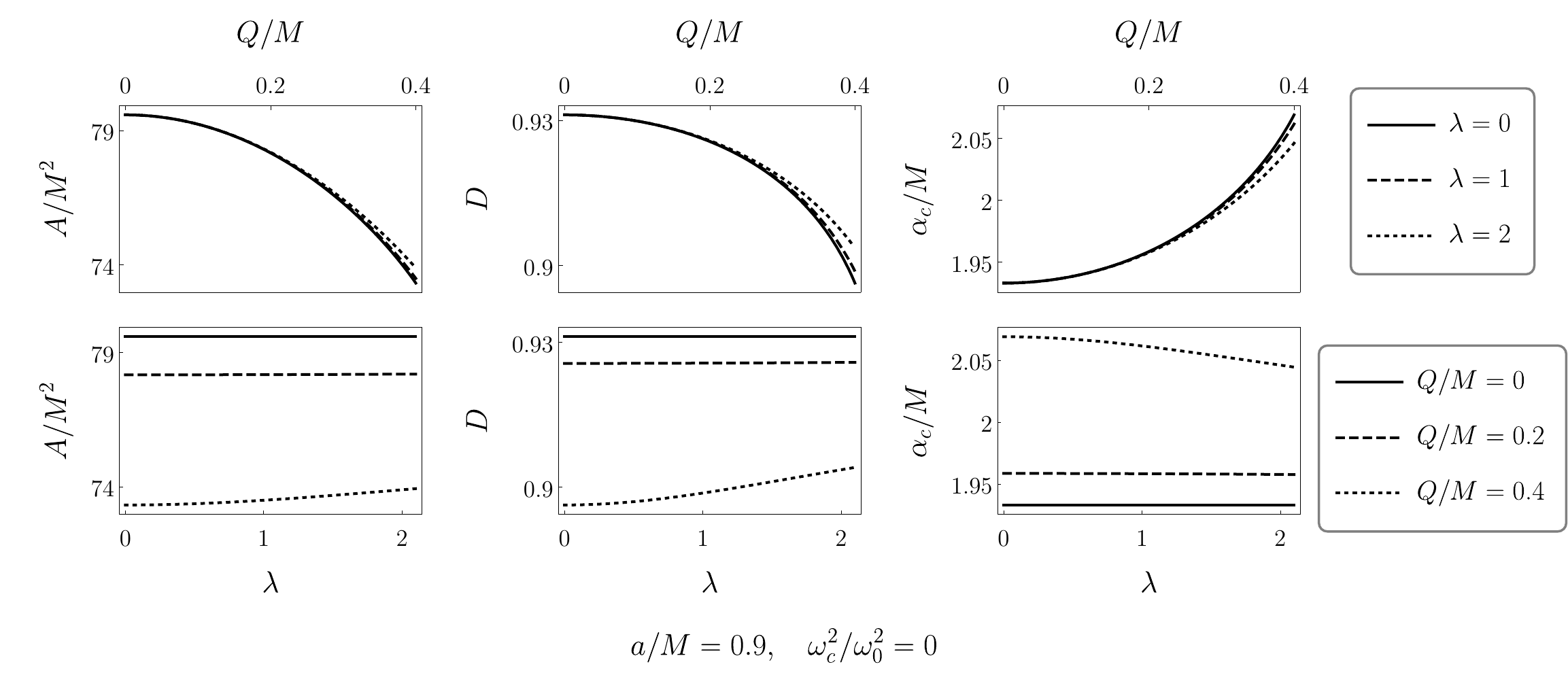}
	\caption{The area $A$, the oblateness $D$, and the horizontal displacement $\alpha_c$ of the shadow for various values of $Q$ and $\lambda$, for photons with $\omega_c^2/\omega_0^2 = 0$. \textit{Top}: the observables as functions of $Q$ for fixed values of $\lambda$. \textit{Bottom}: the observables as functions of $\lambda$ for fixed values of $Q$.}
	\label{fig:obs-inf}
\end{figure}

\begin{figure}
	\centering
	\includegraphics[width=\linewidth]{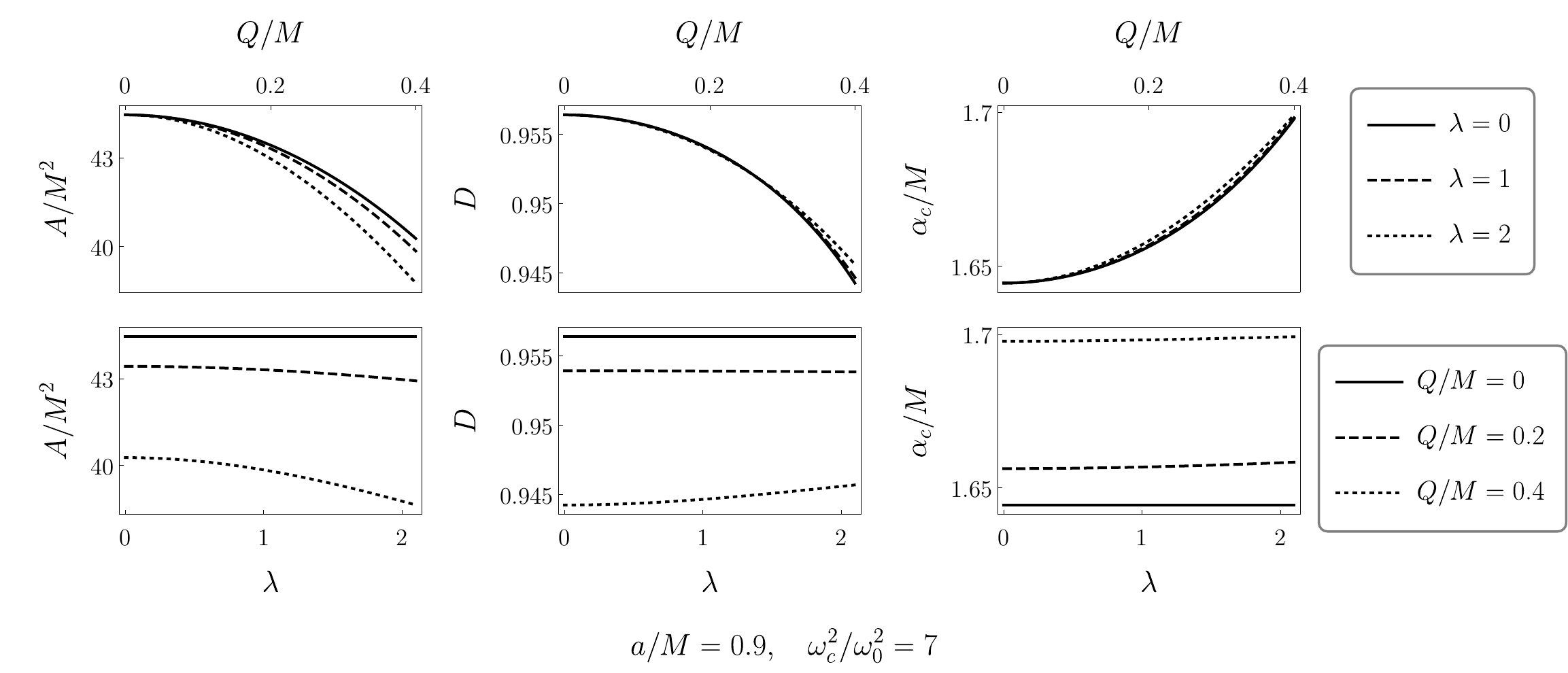}
	\caption{Same as in Fig. \ref{fig:obs-inf}, with $\omega_c^2/\omega_0^2 = 7$.}
	\label{fig:obs7}
\end{figure}

\begin{figure}
	\centering
	\includegraphics[width=\linewidth]{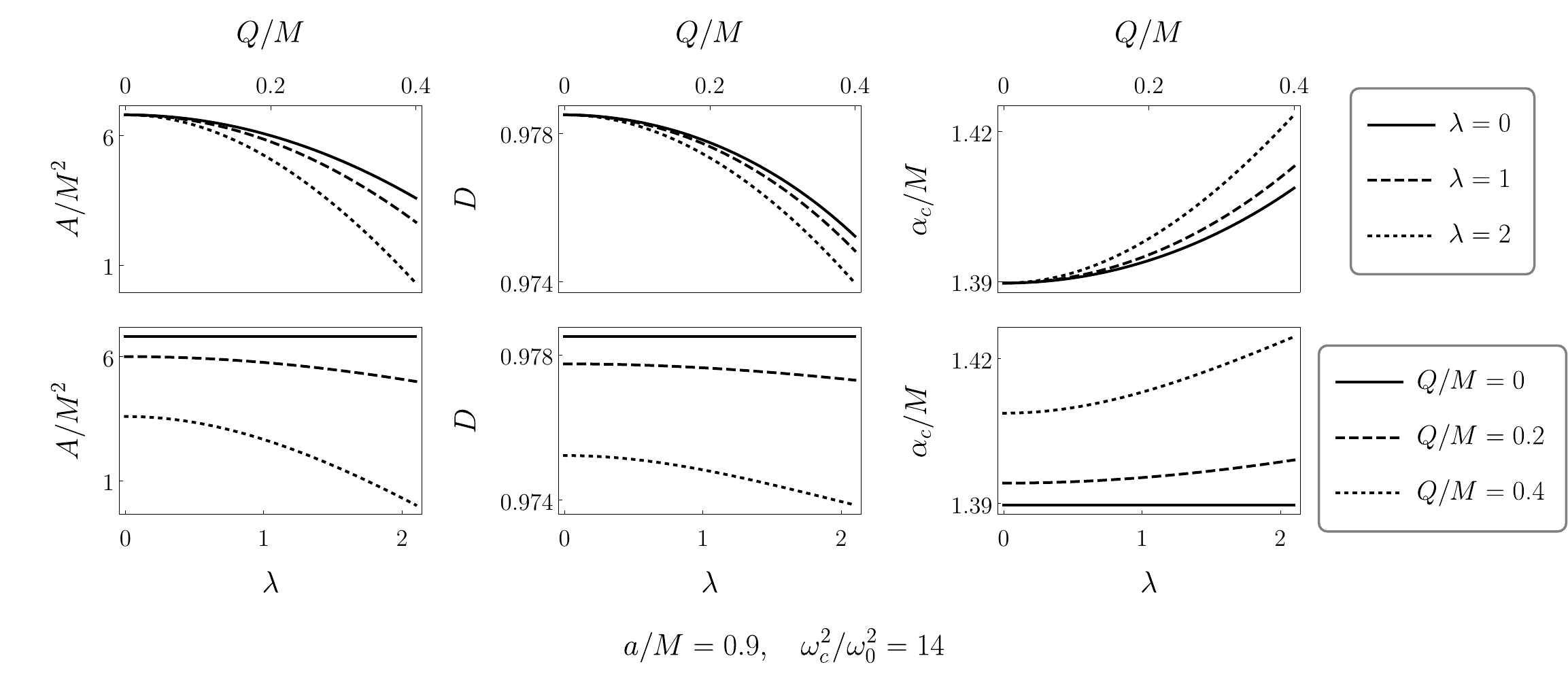}
	\caption{Same as in Fig. \ref{fig:obs-inf}, with $\omega_c^2/\omega_0^2 = 14$.}
	\label{fig:obs14}
\end{figure}

As we have done in previous works \cite{badia20,badia20mg} and following Ref. \cite{kumar20}, for a given black hole shadow we define three geometrical quantities, called observables: the area of the shadow, its oblateness, and the position of the centroid\footnote{Other observables have also been introduced in the literature, e.g. Refs. \cite{parameters}.}. These provide a convenient way of studying how the size, the shape, and the position of the shadow changes with the metric parameters, or of determining them from a hypothetical observation. For a given plasma distribution, the black hole shadow considered here is determined by five parameters: the mass, the spin, and the charge of the black hole, the dilaton coupling, and the observer inclination angle. If two of these are found from astrophysical observations\textemdash for example, the mass, and the spin or the inclination angle\textemdash then the other three parameters can be obtained from the observables, assuming enough experimental precision. Additional observables could also be defined \cite{abdujabbarov15}, but we have chosen these three for simplicity. The area can be calculated by
\begin{equation}
	A = 2 \int \beta\, d\alpha = 2 \int_{r_+}^{r_-} \beta(r) |\alpha'(r)|\, dr,
\end{equation}
with the factor of 2 compensating for the fact that one must choose one sign for $\beta$ in Eq. \eqref{beta}. The oblateness is defined as
\begin{equation}
	D = \frac{\Delta\alpha}{\Delta\beta},
\end{equation}
where $\Delta\alpha$ and $\Delta\beta$ are the diameters of the shadow in the horizontal and vertical directions respectively; it measures the deviation from circularity, with a circular shadow having $D=1$. Finally, the centroid of the shadow is horizontally displaced with respect to the optical axis, with its position given by
\begin{equation}
	\alpha_c = \frac{2}{A} \int \alpha \beta\, d\alpha = \frac{2}{A} \int_{r_+}^{r_-} \alpha(r) \beta(r) |\alpha'(r)|\, dr.
\end{equation}

In order to produce particular examples of shadow images, we have to choose a plasma distribution. Following our previous works \cite{badia21,badia21mg}, we adapt the density profile originally derived for the case of dust at rest at infinity falling into a Kerr black hole \cite{shapiro74}. The electron density in that case goes as $r^{-3/2}$, but a purely radial profile is not of the form \eqref{plasma-sep} and thus does not allow the separation of the equations of motion. We therefore take
\begin{equation}
	\omega_p^2 = \omega_c^2 \frac{M \sqrt{r}}{H}
\end{equation}
to be our plasma distribution, where $\omega_c$ is a constant; it is separable, and goes as $r^{-3/2}$ for $r \gg a$. Another subtlety is that this solution was derived for the Kerr spacetime; we have to assume that using the EMD metric \eqref{metrica-rot} does not significantly alter the plasma distribution.

In Fig. \ref{fig:sombras1} we show the shadow contours for black holes with $a/M = 0.9$ and $Q/M = 0.4$, over a few values of $\lambda$ and the photon frequency $\omega_0$ and as seen by an observer with $\theta_\text{o} = \pi/2$. The value of $Q/M$ was chosen to be close to the extremal value \eqref{q-extremal} over the range of parameters considered, in order that the effect of the electromagnetic and dilaton fields be as large as possible. The most obvious property of the shadow is that its size decreases as the frequency decreases---drastically so for higher values of $\lambda$. This can be traced back to condition \eqref{prohibida}, which dictates the regions where light rays of a given frequency may travel; in particular, it can be seen that as the frequency decreases a forbidden region forms around the poles, as shown in Fig. \ref{fig:regiones}. The deformation and horizontal displacement characteristic of rotating black hole shadows are also present. Decreasing the frequency tends to compensate for these effects, leading to a more circular and centered shadow.

The behavior of the shadow as the parameters of the metric are changed is more easily seen by plotting the three observables $A$, $D$, and $\alpha_c$ as in Figs. \ref{fig:obs-inf}, \ref{fig:obs7}, and \ref{fig:obs14}, where they are shown as functions of $Q/M$ and $\lambda$ for three different frequencies. As in the Kerr-Newman case, the shadow size and its oblateness decrease with $Q/M$, while the horizontal displacement of its centroid increases. In Fig. \ref{fig:obs-inf} it can be seen that in the absence of plasma (which is equivalent to the limit of infinite frequency, i.e. $\omega_c^2/\omega_0^2 = 0$), higher values of $\lambda$ reduce the gravitational effect of the electric charge, bringing the shadow closer to its Kerr shape and size. More explicitly, for a given $Q/M$, the presence of the dilaton makes the shadow become larger and more circular, as well as moving closer to the optical axis. This is expected, since it can be shown that with a fixed value of $Q/M$, the metric approaches to the Kerr metric in the limit that $\lambda$ goes to infinity\footnote{
A change of coordinates bringing the singularity to $r=0$ is needed to show this.
}. However, this deviation is small for the values of $\lambda$ considered, and as the frequency is lowered it is rapidly overshadowed by the presence of the plasma. In Fig. \ref{fig:obs7}, in which $\omega_c^2/\omega_0^2 = 7$, the shadow area and displacement become increasing functions of the coupling for a fixed charge, while the oblateness becomes less sensitive to it. Decreasing the frequency further to $\omega_c^2/\omega_0^2 = 14$, as in Fig. \ref{fig:obs14}, shows that the behavior of all three observables is inverted with respect to the vacuum case: the area, which is already very close to zero, decreases as the coupling becomes stronger. A larger $\lambda$ also leads to a less circular and more displaced shadow, though the variations are very small.

\section{Discussion}\label{sec:conc}

The shadow of a black hole can be a useful probe of the spacetime curvature in the strong gravity region close to the event horizon, and it has gained relevance since the observation of two supermassive black holes by the EHT \cite{eht19,eht22}. In this work, we have explored the shadow of black holes in a generalized Einstein-Maxwell-dilaton theory with a coupling parameter $\lambda$ \cite{ghs91,gibbons88}, relying on the modified Newman-Janis algorithm to produce a rotating counterpart from the static solution \cite{azreg19}. The rotating metric given by the algorithm requires a modified energy-momentum tensor with respect to the original spacetime \cite{azreg19}. A particular case of interest, when $\lambda = 1$, is the metric corresponding to the Kerr-Sen solution of Einstein-Maxwell-dilaton-axion gravity \cite{yazadjiev00,sen92}. We have also included the presence of a very simple plasma model as a way to approximate the chromatic (i.e., frequency-dependent) effects that might be present in the vicinity of an astrophysical black hole. Other important processes like scattering, emission or absorption are not taken into account, which should be included in a more realistic study.

The plasma model has been adapted from the solution corresponding to presureless dust falling into a Kerr black hole \cite{shapiro74}; the specific form of the plasma distribution has been chosen so that the Hamilton-Jacobi equation for light rays is separable, and thus the shadow can be found by using the standard method of finding the spherical photon orbits. Our main results are Eqs. \eqref{pphi-c} and \eqref{K-c}, giving the constants of motion of these orbits, from which the shadow can be easily plotted by using Eqs. \eqref{alfa} and \eqref{beta}, as we have done in Sec. \ref{sec:ex}. In our examples, we have considered fixed values for the rotation parameter $a/M$ and the observer inclination $\theta_\text{o}$, since their effect on the shadow size and shape is already well-known. We have found that, as in other spacetimes with plasma \cite{badia21,badia21mg}, the photon frequency $\omega_0$ is the parameter that has the largest impact on the shadow shape and size: light below a certain frequency cannot approach the black hole and thus produces no shadow, and the area of the shadow decreases rapidly as the frequency approaches its threshold value from above. The qualitative behavior of the shadow when increasing the values of $Q/M$ is similar to the Kerr-Newman case. In addition, for a given value of $Q/M$, the presence of the dilaton has a frequency dependent effect on the shadow. For high frequencies, a higher coupling leads to a larger shadow as compared to the Kerr-Newman case. The shadow is also more circular and is positioned closer to the optical axis. On the other hand, at lower frequencies, below a scale set roughly by the characteristic plasma frequency $\omega_c$, increasing the coupling makes the shadow smaller, more elliptical and less centered.

The results of this work can be applied to the entire range of photon frequencies, from very high frequencies where the effect of the plasma is negligible down to its minimum value, where a forbidden region surrounds the black hole completely and the shadow disappears. In the vicinities of the supermassive black holes Sgr A* and M87* the effects of the plasma start to become relevant at wavelengths greater than around $10\, \mathrm{cm}$ \cite{perlick15}, that is, $\omega_c/2\pi$ is around $3\, \mathrm{GHz}$, while the EHT operates at $1.3\, \mathrm{mm}$, so that $\omega_0/2\pi \approx 230\, \mathrm{GHz}$. The image resolution is not yet sufficient to observe the small change in the shadow area or any of the other two observables produced by the plasma at this low value of $\omega_c^2/\omega_0^2$. The expected variation of the shadow size and shape due to the presence of the electromagnetic and dilaton fields is also within experimental uncertainty, so that it is not yet possible to constrain the values of the dilaton coupling or the electric charge by using the already observed black hole shadows. Discerning the influence of a plasma or the electromagnetic and dilaton fields on the shadow seems to be out of reach for the present and near future facilities.

\section*{Acknowledgments}

This work has been supported by CONICET and Universidad de Buenos Aires.

\end{document}